\documentclass[sigconf]{acmart}
\renewcommand\footnotetextcopyrightpermission[1]{} 

\AtBeginDocument{%
  }

\author{Shuyang Wang}
\affiliation{%
  \institution{Northwestern University}
  \streetaddress{2145 Sheridan Road}
  \city{Evanston}
  \state{IL}
  \country{USA}
  \postcode{60208}
}

\author{Diego Klabjan}
\affiliation{%
  \institution{Northwestern University}
  \streetaddress{2145 Sheridan Road}
  \city{Evanston}
  \state{IL}
  \country{USA}
  \postcode{60208}
}

\acmConference[ ]{ }{ }

\begin{document}

\title{An Ensemble Method of Deep Reinforcement Learning for Automated Cryptocurrency Trading}

\begin{abstract}
  We propose an ensemble method to improve the generalization performance of trading strategies trained by deep reinforcement learning algorithms in a highly stochastic environment of intraday cryptocurrency portfolio trading. We adopt a model selection method that evaluates on multiple validation periods, and propose a novel mixture distribution policy to effectively ensemble the selected models. We provide a distributional view of the out-of-sample performance on granular test periods to demonstrate the robustness of the strategies in evolving market conditions, and retrain the models periodically to address non-stationarity of financial data. Our proposed ensemble method improves the out-of-sample performance compared with the benchmarks of a deep reinforcement learning strategy and a passive investment strategy. 
\end{abstract}



\maketitle

\section{Introduction}
Electronic trading has gained widespread popularity among investors for its low costs and flexibility over the past decades. To facilitate the execution of increasingly large volume of trades on the fully electronic markets, professionals resort to algorithmic trading to automate trading strategies and seek higher efficiency and returns. Many rule-based algorithmic trading strategies still rely on expert knowledge and human supervision. Recent development of deep reinforcement learning (DRL) algorithms demonstrates its potential to learn optimal policies for highly complicated control tasks in stochastic environments\cite{mnih2013playing, silver2017mastering}. The success has inspired researchers to leverage the power of DRL algorithms to find automated trading strategies that are adaptable to evolving market conditions. The attempts so far show some promising progress with evidence of profitable trading strategies in backtesting\cite{chen2019application, jiang2017cryptocurrency, huang2018financial, liu2021finrl, theate2021application}. Nevertheless, the non-stationary and chaotic nature of financial markets poses severe threats to the generalization performance of DRL trading strategies and their deployment in live trading. Our work focuses on the generalization and evaluation of DRL trading strategies on the out-of-sample data. We design an ensemble method and a model selection schema to improve the out-of-sample trading performance measured by the annualized and risk-adjusted returns, and provide a distributional view of the generalization performance of the trained models in changing market conditions.

The generalization of DRL trading strategies remains an active research area and central to their reliable performance in live trading. Numerous attempts have been made to adapt DRL algorithms to accommodate dynamic financial markets, including improvements of feature extraction\cite{wu2020adaptive,  li2019deep}, noise reduction in reward signals\cite{zarkias2019deep}, and ensemble methods\cite{yang2020deep, carta2021multi}. While some existing works simplify the trading environment by allowing only finite actions or a single asset to study improvements in generalization, we explore in the realistic setting that supports continuous actions of trading multiple assets simultaneously while focusing on the out-of-sample trading performance. The challenge for a reliable generalization of DRL strategies is two-fold. The noisy nature of the market hinders the extraction of informative patterns from the historical price and volume data, while the non-stationary property amplifies the overfitting issue in training and leads to unreliable performance on out-of-sample data. On the feature extraction side, we employ the long short-term memory\cite{hochreiter1997long} (LSTM) module to capture the temporal relationships in sequences of historical observations and enrich the features by technical indicators commonly utilized by traders; on the generalization side, we propose a novel mixture distribution policy to effectively ensemble the best performance models from several validation periods. By tracking the moving average of the strategy's performance during training, we retain models that exhibit persistently high performance and robustness to small weight perturbations by recent optimization steps, and avoid those with spurious sudden spikes in performance.  

We break down the out-of-sample data into granular test periods, and retrain and test the model periodically to address the difficulty of training with non-stationary financial time series. We provide a distributional view of the performance on the granular test periods. Although a true probabilistic evaluation of the generalization performance can be hard to achieve due to the scarcity of the price history, the distribution of the performance on granular test periods can still reveal the model performance in multiple historical market conditions. The granular performance evaluation also helps mitigate the critical issue of false positive reporting of trading strategies that achieve impressive overall cumulative returns but are highly overfitted to specific market conditions. Extreme and positive returns in a histogram may overshadow the overall poor performance on other periods, while the extreme and negative returns suggest potential failures in certain market conditions. 

Since the rise of cryptocurrency in recent years, the cryptocurrency market has emerged as an alternative opportunity for investors to diversify their portfolio for its weak correlation with traditional financial markets. Therefore, it is important to understand the dynamics of the cryptocurrency prices to guide investment decisions. However, the cryptocurrency market is characterized by its large volatility and fluctuations. It also lacks fundamental data, for example, earnings per share, making it particularly difficult to conduct valuation and risk factor analyses for cryptocurrencies. Meanwhile, the cryptocurrency market is open around the clock, allowing for a greater potential of applying automated trading strategies. Therefore, we choose cryptocurrency trading as our testing ground to explore reliable automated trading strategies and supplement the literature on intraday cryptocurrency trading using DRL algorithms. We use the FinRL-Meta\cite{liu2022finrl} framework that supports continuous trading actions of a cryptocurrency portfolio as our DRL strategy benchmark, and use a buy and hold strategy as our passive investment benchmark. Our proposed method outperforms both benchmarks in annualized and risk-adjusted returns on 4-year out-of-sample data, showing a promising DRL approach to train reliable automated trading strategies for cryptocurrency portfolio trading. 

Our contribution is as follows.
\begin{itemize}
    \item We propose a novel mixture distribution policy for DRL trading strategies to effectively ensemble the best performance models on multiple validation periods, and achieve high returns on out-of-sample data.
    \item We design a model selection schema that tracks the moving average of the validation performance to select robust models during training.
    \item We provide a distributional view of the out-of-sample performance on granular periods to demonstrate the robustness of DRL strategies in evolving market conditions.
\end{itemize}

\section{Related Work}

The task of portfolio trading can be formulated as a dynamic programming problem where at each time step, the portfolio manager makes a trading decision to maximize the expected return. Many attempts have been made to apply DRL algorithms to find profitable trading strategies. Some works simplify the problem by discretizing the action space\cite{nan2022sentiment, taghian2022learning, kabbani2022deep, hirchoua2021deep, wu2020adaptive, li2020application, li2019deep, chen2019application, liu2018practical}. They use DRL algorithms for a discrete action space where the buy or sell amount of each trade is restricted to a finite set of predefined values. We relax the assumption and study continuous actions of trading a portfolio, which is closer to the realistic trading environment.

Among the works of reinforcement learning method of trading multiple assets with continuous action space, \cite{zhang2020deep} use a reward scaling method that incorporates modern portfolio theory and outperforms the momentum strategies in future contracts market, while we explore trading in the more volatile cryptocurrency market. \cite{guan2021explainable} focus on the explanability of DRL strategies, while our work is focused on the generalization performance. FinRL-Meta\cite{liu2022finrl} proposes a generic framework for trading portfolios in different markets with continuous actions. We use models trained by FinRL-Meta as a benchmark to demonstrate the superiority of our proposed ensemble method.

We supplement the literature on the important issue of reliable generalization of DRL trading strategies in changing market conditions. \cite{kuo2021improving} design a generative adversarial market simulator to improve the generalization of DRL trading strategies. \cite{kochliaridis2023combining} adopt a rule-based mechanism according to the market environment that explicitly controls the actions taken by the trading strategy for reliable generalization. Our work utilizes the power of DRL algorithms without additional expert knowledge. \cite{yang2020deep} design an ensemble method that combines models trained by three DRL algorithms to find optimal trading strategies, which incurs high computational cost of training multiple models. Our work proposes an efficient ensemble method that leverages the rich source of models along one training trajectory. \cite{gort2022deep} take an explicit approach to address the overfitting issue of DRL trading strategies. They add a module of hypothesis testing that rejects overfitted agents to ensure higher chance of good performance on the out-of-sample. Our ensemble method implicitly addresses the overfitting issue by effectively combining models that exhibit good performance on different periods with a novel mixture distribution policy.

Our work is in line with the emerging trend of literature on DRL trading in the cryptocurrency market. Compared with the traditional financial markets, the cryptocurrency market is characterized by its high volatility and fluctuation without reliable fundamental analysis tools. Researchers attempt to find robust DRL strategies amidst these additional challenges for cryptocurrencies. On trading a single cryptocurrency, \cite{sattarov2020recommending} study trading points recommendation by DRL and show profitability; \cite{lucarelli2019deep} apply a double deep Q-learning algorithm to train the model for discrete actions, with rules to stop losses and guarantee gains; \cite{cornalba2022multi} explore using a multi-objective algorithm to improve trading performance for a single asset. On trading a cryptocurrency portfolio, \cite{jiang2017cryptocurrency} train a convolutional neural network to learn portfolio weights as actions; \cite{betancourt2021reinforcement} experiment adding a self-attention layer to the DRL model architecture. We use LSTM as the feature extraction module and focus on the ensemble method for DRL trained models. A DRL approach has also been taken to study cryptocurrency market making using the order book data\cite{sadighian2019deep}.

There have been attempts of using alternative data or approaches to improve upon existing DRL strategies. Some \cite{ye2020reinforcement, koratamaddi2021market, nan2022sentiment, kabbani2022deep} augment the data by market sentiments and show improvement of performance, while we aim to utilize only the technical data to find profitable strategies; \cite{felizardo2022outperforming} propose a supervised learning approach to trading a single asset; \cite{hirchoua2021deep} obtain promising results of a risk curiosity-drive learning framework. Our work demonstrates the power of DRL algorithms in training reliable trading strategies when a novel ensemble method is adopted.

\section{Background}
\subsection{MDP Formulation}
We formulate the problem of trading a portfolio of cryptocurrencies as a Markov Decision Process (MDP). The objective is to maximize the return at the end of the trading period through a sequence of decisions of buying, selling or holding each asset at each time step $t \leq T$. We assume that the market is liquid enough with negligible bid ask spread. The formulation of MDP and the constraints are described as follows, where $D$ is the number of cryptocurrencies. 

\textbf{State space}: At time $t$, the state $s_t = [P_t, I_t, H_t, b_t]$; $P_t $ represents the open, high, low, close prices and trading volume (OHLCV) of time $t$, and $I_t$ consists of additional technical indicators for each cryptocurrency; $H_t \in \mathbb{R}^D, H_t \geq 0$ denotes the holding of each asset and $b_t \geq 0$ denotes the balance in USD on the account. 

\textbf{Action Space}: At time $t$, the agent takes an action $a_t \in \mathbb{R}^D$ that changes the holding of the cryptocurrency $d$ by $a_{t,d}$, subject to constraints of non-negative holdings and non-negative balance on the account. The agent takes actions according to a policy $\pi_t = \pi(s_t)$ that maps the states to the probability of selecting the actions. Additionally, the trade size is scaled by a predefined value $hmax_d$ such that $|a_{t,d}| \leq hmax_d$ for all $d \in [D]$. 

\textbf{Transition}: $\text{Pr}(s_{t+1} | s_{t}, a_{t})$ denotes the probability of transitioning from state $s_t$ to $s_{t+1}$ by taking action $a_t$. The trading environment determines $P_{t+1}$ and $I_{t+1}$, and the action modifies $H_{t+1}$. The balance $b_{t+1}$ is adjusted after executing the trades at close prices at time $t$. A real number reward $r_t$ is collected after the transition.

\subsection{Technical Indicators}
We use the following technical indicators to supplement the OHLCV data as additional features in the states. Technical indicators are widely adopted by traders to detect trend and momentum in financial time series and provide important guidance for their trading decisions. 
\begin{itemize}
    \item Simple moving average (SMA) tracks the price trend using the arithmetic average of the prices at $p$ recent time steps. 
    \item Relative strength index (RSI) is a momentum oscillator in the range of $[0, 100]$ that is used to validate the price trends. An RSI below 30 signals an oversold condition while an RSI above 70 is usually considered an overbought condition.
    \item Commodity channel index (CCI) is an unbounded oscillator that measures the difference between the current price and the average of past prices in a typical look-back window of 20 time periods. CCI increasing above 100 can indicate an emerging bullish trend while CCI dropping below -100 can signal a new bearish trend.
    \item Moving average convergence/divergence (MACD) is a momentum indicator that calculates the difference between the exponential moving averages of 26 periods and 12 periods.
    \item Average true range (ATR) is a non-negative volatility indicator that measures the average price movement range. A larger value indicates a larger volatility.
    \item Average directional index (ADX) is a trend strength indicator in the range of $[0, 100]$. It signals the presence of a strong trend in either direction when ADX is above 25. 
\end{itemize}

\subsection{Long Short-Term Memory Network}
The LSTM architecture is designed to address the vanishing and exploding gradient issues of recurrent neural networks\cite{hochreiter1997long}. The model uses gates to control the read and write of the memory and state cells, so that the model is capable of learning long-term dependencies in a sequence. We use LSTM to extract the patterns in a sequence of observations.

\subsection{Proximal Policy Optimization}
We use the Proximal Policy Optimization (PPO) algorithm\cite{schulman2017proximal} to train the DRL agent to maximize the discounted return $R = \mathbb{E}\sum_t \gamma^t r_t$, where the reward $r_t$ is the change in the portfolio value at time step $t$ of an episode and $\gamma \leq 1$ is the discount factor. An episode extends from the beginning to the end of the trading period. PPO is an actor-critic algorithm where the actor generates a policy compatible with continuous action spaces, and the critic learns the value of the current state. PPO is built on the trust region policy optimization algorithm and uses a clipped surrogate objective function to encourage conservative policy updates. It achieves state-of-the-art or comparable performance while enjoying the benefit of a relatively easy implementation. 

\section{Methodology}
\subsection{Model Selection}
We propose a model selection schema that evaluates on multiple validation periods during training by rolling out independent validation episodes, and tracks the moving average of the validation returns. We randomly select $K$ in-sample periods of the same length $T_{test}$ as test periods as the validation periods. In-sample validation may exacerbate overfitting in supervised learning, but for reinforcement learning, the issue can be less severe. The state vectors differ in the holdings of assets and account balance for the same period visited in training and validation stages, which leads to potentially different policy outputs and subsequent trajectories. Given the non-stationary property of financial time series, multiple validation periods can provide a more comprehensive representation of market conditions. The validation returns evaluated on periods of the same length as testing time horizons also provide a more accurate assessment of the model performance, compared to a single return evaluated over the entire training period. Additionally, we track the moving average of validation returns over several past iterations. By selecting models with the highest smoothed return, the model selection method implicitly favors models that are robust to weight perturbations by optimization steps and demonstrate consistently high profitability.

\subsection{Ensemble Policy}
We model the stochastic policy by a $\tanh$ transformed Gaussian distribution $TanhGaussian(\mu, \Sigma)$ where $\Sigma$ is a $D\times D$ diagonal matrix with $\sigma^2_d$ on the diagonal. The LSTM model outputs $\mu_d, \sigma_d$ for each $d \in [D]$. Sampling from a $\tanh$ transformed Gaussian distribution can be interpreted as sampling an auxiliary value $\Tilde{a}_d$ from a Gaussian distribution $\mathcal{N}(\mu_d, \sigma^2_d)$, applying $\tanh$ to map the sampled value to the interval of $(-1, 1)$, and scaling the result by a predefined factor $hmax_d$. The sampled action satisfies the constraints $|a_d| < hmax_d$.

For the inference on out-of-sample periods, we propose an ensemble policy using a novel mixture of $\tanh$ transformed Gaussian distributions to effectively combine the policies from the $K$ sets of model weights at the epochs where the model achieves the highest smoothed validation performance. At each time step $t$ of the test episode, the agent takes an action according to the ensemble policy $\pi_t(a) = \frac{1}{K} \sum_{k=1}^{K} TanhGaussian_k (a | \mu_k(s_t),\Sigma_k(s_t))$.

Models that achieve superior performance on more validation periods demonstrate better capacity to maintain profitability in different market conditions, and have higher weights in the mixture of equal weighted individual distributions. While for models that are overfitted, their individual effects can be reduced by the ensemble policy. Given a specific environment observation, actions that are favored by more models, including the ones overfitted to their corresponding validation periods, represent an approximate consensus among all policies. The ensemble policy can be viewed as a continuous counterpart to the majority vote in discrete cases. With little additional computational overhead, we efficiently leverage the various market conditions in the history and the models along the training trajectory to enhance the generalization ability of DRL trading strategies.

\subsection{Evaluation on Granular Test Periods}
We break down the out-of-sample data into granular periods and evaluate the model performance on each test period to gain a distributional understanding of the generalization performance in evolving market conditions. We retrain the model periodically using a rolling window of the data preceding the test periods to adapt the model to recent market conditions. By comparing the quantiles of the test performance with those of the market performance, we assess the robustness of DRL strategies behind an overshadowing return on the entire test data. Extreme values in the distribution can indicate either severe failures in certain conditions or rare successes that obscure overall poor performance. The identification of the extreme cases also helps mitigate the false positive reporting of trading strategies that are highly overfitted to specific market conditions despite an overall high return, which is a crucial step before the deployment of trading strategies in live trading.  

\section{Computational Experiments}
\subsection{Data Processing}
We use historical hourly OHLCV data from the online cryptocurrency exchange Kraken for the period from 01/01/2018 to 06/30/2022. We consider a portfolio constituted by USD and 5 largest cryptocurrencies by market capitalization in the month of December 2017, namely Bitcoin (XBT), Ethereum (ETH), Bitcoin Cash (BCH), Ripple (XRP) and Litecoin (LTC). For each cryptocurrency, we detrend the time series by taking the percentage change of the raw OHLCV data, to avoid serious failures in generalization when the absolute price range enters a different regime in out-of-sample data. The price trend and momentum information is extracted and preserved by the 7 technical indicators, including $SMA_{30}$ and $SMA_{60}$ of close price for 30 and 60 periods, $RSI$, $CCI$, $MACD$, $ADX$ and $ATR$.

For each time step $t$, the price and technical features of the 5 cryptocurrencies form a vector $f_t \in \mathbb{R}^{60}$. The features are concatenated with the holdings of the assets and the balance in USD to form a vector $\Tilde{s}_t \in \mathbb{R}^{66}$. Since the financial time series is generally considered non-Markovian, we define the states $s_t \in \mathbb{R}^{12\times 66}$ as a sequence $[\Tilde{s}_{t}, \Tilde{s}_{t-1}, ..., \Tilde{s}_{t-11}]$ and assume that the Markovian property is satisfied for the constructed states $s_t$.

We apply feature-wise normalization to the states $s_t$. We standardize the percentage changes of OHLCV, $SMA_{30}$, $SMA_{60}$, $CCI$, $MACD$ and $ATR$. $RSI$ and $ADX \in [0, 100]$ are linearly mapped to the interval of unit length centered at 0. For holdings and balance, we apply min-max normalization with a predefined maximum value for all $t$ based on the initial portfolio value and initial prices.

\subsection{Model Architecture} The inputs $s_t$ are of dimension $12 \times 66$. We use a fully connected layer followed by a 2-layer LSTM with hidden dimension of 64 for each layer to learn the representation of the sequential data. The sequential feature extraction module is shared by the policy and the value networks. Th policy network uses fully connected layers to output vectors $\mu, \sigma$ as parameters of the stochastic policy, and the value network uses a fully connected layer to output the value of the state. The network architecture is detailed in Table~\ref{tab:nnarch} and the weights are initialized by the Xavier initialization method.

\subsection{Training} We use the 4-year data from 07/01/2018 t0 06/30/2022 for backtesting. The test data is split into 48 monthly periods. For each out-of-sample period, we use the preceding 6-month of data to train for a total of 400 episodes rolled out by 4 agents in parallel, where each episode starts with 1 million USD on the account and zero holdings of cryptocurrencies. The episodes vary only by stochasticity of the policy. We sample 9 weekly periods within the training period as validation sets and roll out validation episodes every 2 epochs of training. 
The reward is calculated by the change of the portfolio value after each action and is normalized by the running maximum of past absolute values of rewards. The action is scaled by a predefined value $hmax$ for BTC and by $hmax$ multiplied by the initial price ratio to BTC for other cryptocurrencies. Based on the immediate rewards and the value function outputs, we apply generalized advantage estimation\cite{schulman2015high} to calculate the estimated advantages $\hat{A}_t$ with a discount factor $\gamma=0.99$ for the rewards and an exponential weight discount factor $\lambda=0.95$ for the extended advantage estimators. The objective function for the PPO algorithm consists of the clipped surrogate policy objective, the mean squared error of the value function loss and an entropy term to control exploration and exploitation, weighted by hyperparameters $c_1, c_2$ and $c_3$. The loss is calculated in batches and minimized by the Adam\cite{kingma2014adam} optimizer with a linear decaying learning rate schedule. The set of hyperparameters used is summarized in Table~\ref{tab:hparams}.

\begin{table*}
\caption{Hyperparameters}
\label{tab:hparams}
    \begin{tabular}{cccccccccc}
    \toprule
    input sequence length & episodes & learning rate & batch size & hmax & $\gamma$ & $\lambda$ &  $c_1$ & $c_2$ & $c_3$ \\
    \midrule
    12 & 400 &  $5 \cdot 10^{-6}$ & 6,000 & 70 & 0.99 & 0.95 & 100 & -1 & -1 \\
    \bottomrule
    \end{tabular}

\end{table*}

\begin{table*}
\caption{Neural Network Architecture}
\label{tab:nnarch}
\begin{tabular}{cccc}
\toprule
Layer & Description & Output shape & Activation \\
\midrule
Fully Connected & 16 hidden units & (12, 16) & sigmoid \\
LSTM & 2 layers of 64 hidden units & 64 & tanh \\
Bacth Normalization & - & - & - \\ 
Fully Connected & policy network: mean & $5 $ & tanh \\
Fully Connected & policy network: standard deviation & 5 &  sigmoid \\
Fully Connected & value network & 1 & linear \\
\bottomrule
\end{tabular}
\end{table*}

\subsection{Metrics} After one iteration of training on a 6-month rolling window, we conduct backtesting of the DRL trading strategies on the weekly periods in the subsequent one-month out-of-sample data. In total, we backtest on 208 weekly periods of the 4-year test data. For each test period, we calculate the annualized and cumulative returns to directly measure the profitability of the strategy. We also assess the risk-adjusted returns of the trading strategy by the Sortino and Sharpe ratios, where the Sortino ratio measures the return adjusted by downside risks while the Sharpe ratio factors in both upside and downside risks. We use maximum drawdown as an indicator of the downside risk and the standard deviation of hourly returns as a proxy of volatility. The annualized return serves as the primary evaluation metric for the trading performance over a given period, while the Sortino ratio is utilized as an important supplementary metric where the undesired downside risks are factored in.

\subsection{Benchmarks}
\textbf{FinRL-Meta:} The framework allows for continuous action spaces and intraday trading of a cryptocurrency portfolio. We include the same technical indicators in addition to the hourly OHLCV data as the input price features of the five cryptocurrencies. The FinRL model is trained by PPO algorithm with a 6-month rolling window for 400 episodes using learning rate of $5 \cdot 10^{-6}$ and scale of the action of $hmax_d=70$ for all $d$. Similarly, we retrain the FinRL model periodically and backtest on 208 weeks of the 4-year test period. These hyperparameters used have been tuned. 

\textbf{Buy and hold:} We use the performance of the buy and hold strategy to benchmark our DRL trading strategy with the market performance. For each of the 208 test weeks, the initial fund of 1 million USD is allocated equally among the five cryptocurrencies at the beginning of the period, and the positions are held until the end of the period. The strategy can be regarded as a passive investment strategy that rebalances on a weekly basis to maintain equal allocation of funds among portfolio constituents. Its performance on the 4-year test data closely tracks the market performance of the five cryptocurrencies.

\section{Results} 
\subsection{Backtesting Results}
We present the out-of-sample backtesting results on the entire test period of 4-year in Table~\ref{tab:backtest} for the ensemble method and the benchmarks. Our ensemble method shows superior profitability in both returns and risk-adjusted returns. The ensemble policy method outperforms the FinRL model in the annualized return by 47\% and the buy and hold strategy by 23\%. FinRL achieves an overall low risk level with the lowest maximum drawdown and volatility, but our ensemble method still outperforms FinRL in the Sortino ratio by 18\% and the buy and hold strategy by 29\%. 

\begin{figure*}
    \centering
    \includegraphics[width=0.85\textwidth]{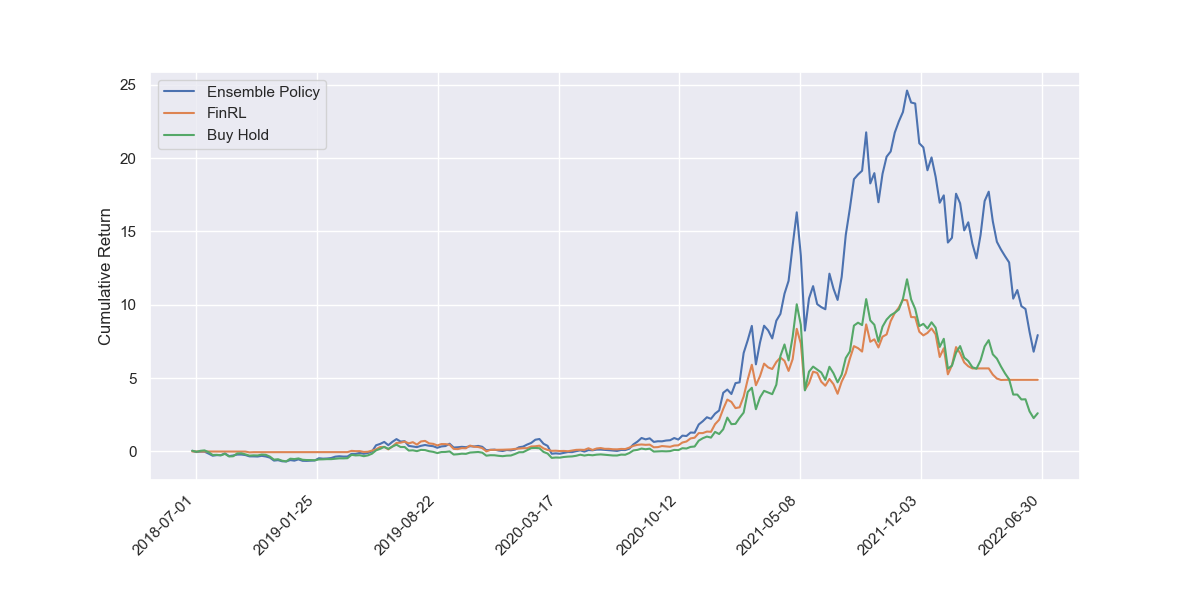}
    \caption{Cumulative returns on 4-year out-of-sample period}
    \label{fig:cumret}
\end{figure*}

The cumulative returns for the 4-year test period are presented in Figure~\ref{fig:cumret}. We observe that the ensemble method follows the market trend in the early stage before 2020. The strategy takes a notable advantage of the bullish trend during the cryptocurrency market rally from 2020 to 2021, which results in cumulative returns of up to 25 times until the market experienced a downturn at the end of 2021. The performance is affected by the market crash reflected in the 69.53\% maximum drawdown, but it is not as severe as the 74.30\% drawdown of the buy and hold strategy. The ensemble method excels at exploiting the upward trend in the market, and shows limited protection during market crashes.

\begin{table}
\caption{Backtesting summary}
\label{tab:backtest}
\begin{tabular}{ cccc } 
\toprule
 & Ensemble policy & FinRL & Buy-hold \\
 \midrule
Annualized return & \textbf{0.9319} & 0.6320  &  0.7571   \\ 
Cumulative return & \textbf{7.9361} & 4.8851  &  2.6088  \\ 
Sortino & \textbf{1.6218}   & 1.3762  &  1.2560   \\ 
Sharpe & \textbf{1.0724} & 1.0401  &  0.8119   \\ 
Max drawdown &  0.6953  & \textbf{0.4818} &   0.7430   \\
Volatility & 0.8690 & \textbf{0.6076} &   0.9325   \\
\bottomrule
\end{tabular}
\end{table}

\subsection{A Distributional View of Returns}

\begin{table}
\caption{Quantiles of weekly returns distribution}
\label{tab:quant}
    \begin{tabular}{cccc}
    \toprule
    Quantile & Ensemble Policy & FinRL & Buy-hold \\
    \midrule
    100\%       &      0.4254 & 0.2825  &  \textbf{0.5630} \\
    90\%       &      \textbf{0.1480} &  0.1243  &  0.1470 \\
    80\%       &      \textbf{0.1085} & 0.0693  &  0.0958 \\
    70\%       &      0.0644  & 0.0240 &  \textbf{0.0646} \\
    60\%       &      \textbf{0.0411} & 0.0061   & 0.0313 \\
    50\%      &      \textbf{0.0210} & 0.0000   & 0.0144 \\
    40\%       &     -0.0112  & \textbf{-0.0000} &  -0.0118 \\
    30\%       &    -0.0307 & \textbf{-0.0066} &  -0.0368 \\
    20\%     &      -0.0647 & \textbf{-0.0284} & -0.0705 \\
    10\%      &     -0.1218 & \textbf{-0.0835} &  -0.1274 \\
    0\%     &      -0.3882  & \textbf{-0.3774} & -0.4633 \\
\bottomrule
    \end{tabular}
\end{table}

In addition to a qualitative description of a strategy's performance from the cumulative returns plots, we take a distributional view of the out-of-sample performance on granular test periods. We present the quantiles of the 208 weekly returns in Table~\ref{tab:quant}. The 100\% quantile is lower than the annualized return in Table~\ref{tab:backtest} since the raw weekly returns without annualization are used to calculate these quantiles. By examining the extreme values, we observe that returns of our proposed method have less extreme returns on both positive and negative sides comparing with the buy and hold returns. The ensemble method shows robust performance relative to the market performance. Returns of our proposed strategy also dominate at most quantile levels at or above 50\% except being slightly behind the buy and hold strategy at 70\%, showing a high concentration on the large positive returns comparing with the benchmarks. While the ensemble method underperforms the FinRL model in preventing losses, it still outperforms the buy and hold strategy at all quantile levels at or below 60\%. Our ensemble method also has the highest median return among the three strategies. The distributional view shows that the returns of the ensemble method concentrate at higher values compared with both the FinRL and the buy and hold strategy, and demonstrates some extent of downside risk control compared to more cases of large losses incurred by the passive investment strategy. The results corroborate the finding that the ensemble method achieves much higher positive returns in a bullish market from the cumulative returns plot.

\begin{figure}
     \centering
         \centering
         \includegraphics[width=0.45\textwidth]{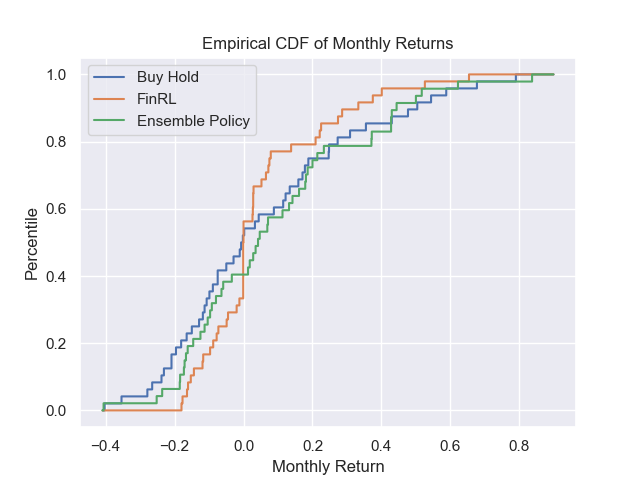}
         \caption{Empirical CDF of monthly returns}
         \label{fig:ecdfret}
\end{figure}

\begin{figure}
         \centering
         \includegraphics[width=0.45\textwidth]{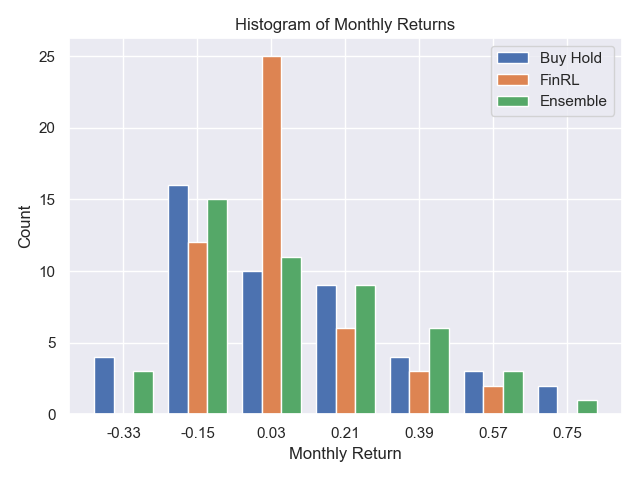}
         \caption{Histograms of monthly returns}
         \label{fig:histret}
\end{figure}

We can also take a distributional view of the performance of the 48 models trained on the rolling window periods. Since the model is retrained every month, we aggregate the weekly returns obtained by the same model trained on the preceding 6-month rolling window to calculate 48 monthly returns. The distribution of the monthly returns reflect the robustness of the performance of the trained model to different training processes. We present the empirical cumulative density function (eCDF) plots for our proposed method and the two benchmarks in Figure~\ref{fig:ecdfret}. The monthly returns of our proposed models dominate the buy and hold returns at all levels for negative returns, and dominate the FinRL model at all levels for positive returns. As we can see from the histograms of the monthly returns in Figure~\ref{fig:histret}, the FinRL model lacks the capacity to produce high returns and the returns are highly concentrated on the insignificant values around zero, while the the buy and hold returns take more extreme values on both positive and negative sides. The ensemble method strikes a balance that avoids severe negative returns but also generates more significant positive returns. 

\subsection{Effectiveness of the Ensemble}

\begin{table}
\caption{Performance of model at last epoch}
\label{tab:lastepoch}
\begin{tabular}{ ccc } 
\toprule
 & Final epoch  & Ensemble policy \\ 
 \midrule
Annualized return &  0.8092  & \textbf{0.9319} \\ 
Cumulative return &  3.7503  & \textbf{7.9361} \\ 
Sortino & 1.3804 & \textbf{1.6218} \\ 
Sharpe & 0.8865   & \textbf{1.0724} \\ 
Max drawdown &  0.8005  &  \textbf{0.6953} \\
Volatility & 0.9128   & \textbf{0.8690} \\
\bottomrule
\end{tabular}
\end{table}

We present the evidence of the effectiveness of the ensemble policy by comparing with the model at the last epoch of training and with the average performance of individual models. The model at the final epoch of training is susceptible to overfitting the training data, and there is little guarantee that the market will exhibit similar trends in the subsequent month. Therefore, in the event of a regime shift, such models may cause serious failures on out-of-sample data. Our ensemble policy accounts for the consensus among different models saved during training to overcome the overfitting issue of the last epoch model. As shown in Table~\ref{tab:lastepoch}, the ensemble model outperforms the final epoch model in all metrics considered, including an improvement of 15\% in annualized return, 17\% in Sortino ratio, and 13\% less in maximum drawdown. The results provide evidence that the ensemble method mitigates the overfitting issue at the end of training. 

\begin{table}
\caption{Performance of individual models}
\label{tab:ind-perf}
\begin{tabular}{ cccc } 
\toprule
   &         Average &   Median & Standard deviation\\
\midrule
Annualized return &  0.8651   & 0.8698 & 0.0802  \\
Cumulative Return  & 5.8103  & 5.9666 & 1.7859  \\
Sortino     &        1.5485   &1.5864  & 0.1534\\
Sharpe     &         0.9713 & 0.9877  & 0.0714   \\
Max Drawdown     &   0.7662   & 0.7756 & 0.0257\\
Volatility       &   0.8895  & 0.8938 & 0.0224   \\
  \bottomrule
\end{tabular}
\end{table}

To show that our proposed method effectively ensembles the individual models, we verify that the ensemble policy produces better results than the expectation of the performance of individual constituent models. We conclude from the aggregated results of the individual models in Table~\ref{tab:ind-perf} that the ensemble method achieves better performance in all metrics considered comparing with the mean and median values of the performance of individual models. The results support the effectiveness of the ensemble policy compared to the average of individual models selected based on a single validation period.

\section{Conclusion and Future Work}
In this work, we focus on improving the out-of-sample performance of DRL trading strategies using an ensemble method. We propose a novel mixture distribution policy that effectively ensembles multiple models selected by the smoothed validation performance on different periods during training. Our method efficiently leverages the historical market conditions and the models along the optimization trajectory to mitigate the overfitting issue for the model at the last epoch of training. Compared with the DRL strategy benchmark FinRL-Meta and the buy and hold strategy, our proposed method demonstrates a significant improvement in out-of-sample returns. We evaluate the model performance on granular test periods and present a distributional view of the returns. The distribution of the returns shows robustness of our ensemble strategy compared with market performance. Our returns show a higher concentration on the large returns compared with the FinRL returns, and also exhibit less extreme positive and negative returns compared with the buy and hold strategy.

The results open up a promising direction to design robust DRL trading strategies using a novel ensemble policy. Future work can incorporate the covariance matrix to the $\tanh$ transformed Gaussian policy to capture the relationships among the portfolio constituents, where the covariance can be learned by the policy network or estimated by historical asset returns. We present a preliminary attempt of learning a covariance matrix for the $\tanh$ transformed Gaussian policy by the policy network. In addition to the outputs of $\mu, \sigma \in \mathbb{R}^D$, the policy network outputs a vector $v \in \mathbb{R}^{1/2D(D-1)}$. We then form a $D\times D$ lower triangular matrix $L$ with $\sigma$ on the diagonal and the elements of $v$ in the off diagonal entries, and construct a positive semi-definite matrix $Cov = LL^T$ as the covariance matrix for the policy. We train the models with the modified policy network and backtest on the 208 weekly periods. The annualized return is 0.6832 and the Sortino ratio is 1.2620. The modified model underperforms our proposed model, though it outperforms the buy and hold strategy in the Sortino ratio by a small margin. Future work can improve the learning of the covariance matrix using the covariance of historical asset returns as a warm start. 

Alternatively, future work can relax some assumptions and study in a more general setting. For example, we can include the constraint of transaction costs in the model and modify the objective to be maximizing of the expected return while minimizing the transaction costs. We can also relax the assumption of non-negative holdings to allow both long and short positions and explore hedging strategies by DRL algorithms.

\bibliographystyle{ACM-Reference-Format}
\bibliography{references}

\end{document}